\documentclass[11pt]{article}
\usepackage{amsmath,amsthm,latexsym,amssymb,amsfonts,amsthm}
\usepackage{graphicx, color, calc, dsfont}
\usepackage[list=true, font=large, labelfont=bf,
labelformat=brace, position=top]{subcaption}

\makeatletter \@addtoreset{equation}{section} \makeatother

\newtheorem{Definition}{Definition}[section]
\newtheorem{Lemma}{Lemma}[section]

\title{Non-convex 4d polytopes in Spin Foam Models}
\author{Benjamin Bahr$^1$\\
\small $^1$ II Institute for Theoretical Physics\\
\small University of Hamburg\\
\small  Luruper Chaussee 149\\
\small 22761 Hamburg, Germany
 }
\date{}

\begin{document}

\maketitle

\begin{abstract}
In this article we consider non-convex $4d$ polytopes in $\mathbb{R}^4$. The paper consist of two parts: Firstly, we extend the proof of the formula for the $4d$ volume in terms of $2d$ face bivectors and boundary graph crossings from convex to non-convex polytopes. Secondly, we consider the EPRL-FK spin foam model, and demonstrate that there exists boundary data which leads to non-convex $4d$ polytopes in the asymptotic analysis of the vertex amplitude. 
\end{abstract}

\section{Introduction}

$4d$ polytopes play an important role in the geometric interpretation of the Euclidean signature EPRL-FK spin foam model for quantum gravity, as they appear in a certain asymptotic regime of the quantum amplitude of this and also related models. \cite{Barrett:1997gw, Engle:2007wy, Freidel:2007py, Baratin:2008du, Bianchi:2010gc, Baratin:2011hp} This has long been seen as a hint that these models are quantum theories of fluctuating $4$-geometries. \cite{Baratin:2008du, Barrett:2009gg, Conrady:2008mk}

However, there are several open questions. Firstly, it has been observed that, for every geometry, there is also the one with reversed orientation\footnote{Under this change the Einstein-Hilbert action $S_{EH}$ reverses its sign, so instead of $\exp(iS_{EH})$ one has $\exp(iS_{EH})+\exp(-iS_{EH})$, which is why this has been dubbed the ``cosine problem''. See \cite{Engle:2011un, Engle:2015mra} for further discussions. }. Secondly, it has been shown that several of the asymptotic geometries do not seem to have an interpretation in terms of $4d$ flat polytopes at all, indeed not even any metric geometries\footnote{These so-called ``twisted geometries'' have been investigated in detail, see e.g.~\cite{Freidel:2010aq, Freidel:2013bfa}.}. Recently, even further non-metric fluctuations\footnote{These only occur in vertices which are more complicated than the 4-simplex, i.e.~in the KKL-extension of the model \cite{Kaminski:2009fm}. They appear to be a consequence of an insufficient implementation of the volume simplicity constraint, see \cite{Bahr:2015gxa, Dona:2017dvf,  Belov:2017who, Bahr:2017ajs}.} have been discovered, which do not seem to be suppressed in the asymptotic limit. All of these geometries can be apparently avoided by carefully choosing the boundary data. 

In this article we report on yet another property of the amplitude. Namely, even if the boundary data is chosen carefully so as to allow only metric geometries in the asymptotic limit, there are cases in which two inequivalent $4d$ geometries arise from the same boundary data. Both of these appear in the asymptotic limit of the amplitude. In our examples, these arise due to the fact that there are examples of a convex and a non-convex polytope having the same intrinsic 3d boundary geometry. This shows that non-convex polytopes arise as geometries in  spin foam models.\\

This article consists of two parts:\\

In the first part, we consider a generalised definition of polytope in $\mathbb{R}^4$, which allows for convex and non-convex ones. We give a definition in section \ref{Sec:Polytopes} and prove a formula for their volume in terms of their bivectors and the crossings in their boundary graphs in sections, \ref{Sec:SubdividingPolytopes} and \ref{Sec:BoundaryGraphs}. This formula has been proven for the convex case in \cite{Bahr:2018pfj}, and has been used to investigate the issue with the volume-simplicity constraint \cite{Bahr:2017ajs}. The generalisation of the formula to the non-convex case is therefore also desirable for the study of the models.\\

In the second part, in section \ref{Sec:InSFM}, we consider an explicit example of boundary data which allows for more than one $4d$ polytope in its asymptotic limit. The volume formula from the previous part will help us in the geometric interpretation of these geometries. Although the construction is within a specific example, it can be easily generalised, and it shows that non-convex geometries are ubiquitous in the EPRL-FK model.\\

We will discuss the findings in section \ref{Sec:Summary}.

\section{Polytopes}\label{Sec:Polytopes}

A convex polytope can easily be represented by the intersection of a collection of half-spaces. The definition of a non-convex polytope is more subtle, and there are several, not quite equivalent definitions available. Intuitively, we work with p.l.~linear subsets $P\subset \mathbb{R}^4$, which come with a choice of triangulation, as well as a partition of the boundary into sub-polytopes. We use the notion of abstract polytopes from \cite{Polytopes}.

\begin{Definition}
A $4d$ polytope is a triple $(P,\Sigma, \tau)$, consisting of the following:
\begin{itemize}
\item A p.l.~subset $P\subset \mathbb{R}^4$ which is a combinatorial $4$-ball,
\item a purely $4$-dimensional simplicial complex $\Sigma$ such that there is a p.l.~homeomorphism $\phi:|\Sigma|\to P$,
\item a map $\tau:\Pi\to \text{Sub}(\Sigma)$ from an abstract polytope $\Pi$ to the subcomplexes of $\Sigma$ with the following properties:
\begin{enumerate}
\item The maximal face of $\Pi$ is a 4-face, i.e.~$\Pi$ is 4-dimensional.
\item One has $\tau(F_{-1})=\emptyset$ and $\tau(F_4)=\Sigma$.
\item For every $k$-face $f$, $\tau(f)$ is a combinatorial $k$-ball.
\item For any two faces $f<g$, $\tau(f)$ is a subcomplex of the boundary $\partial\tau(g)$ of $\tau(g)$.
\item For a $k$-face $g$, any $k-1$-simplex $\sigma$ in the boundary of $\tau(g)$ is in the image of exactly one $f<g$.
\end{enumerate}
\end{itemize}
The polytope is called \emph{nice} if:
\begin{itemize}
\item For a $3$-face, $\phi(|\tau(f)|)$ lies in a proper $3d$ hyperplane of $\mathbb{R}^4$.
\end{itemize}
\noindent We recall that for a combinatorial $k$-ball $B$, the subcomplex $\partial B$ is well-defined, is generated by all $k-1$-simplices of type 1 (i.e.~those which are part of only one $k$-simplex), and is homeomorphic to a $k-1$-sphere. Also note that $\phi$ restricted to $|\partial \Sigma|$ provides a triangulation of $\partial P$. 
\end{Definition}

\begin{figure}[hbt!]
\begin{center}
\includegraphics[scale=0.5]{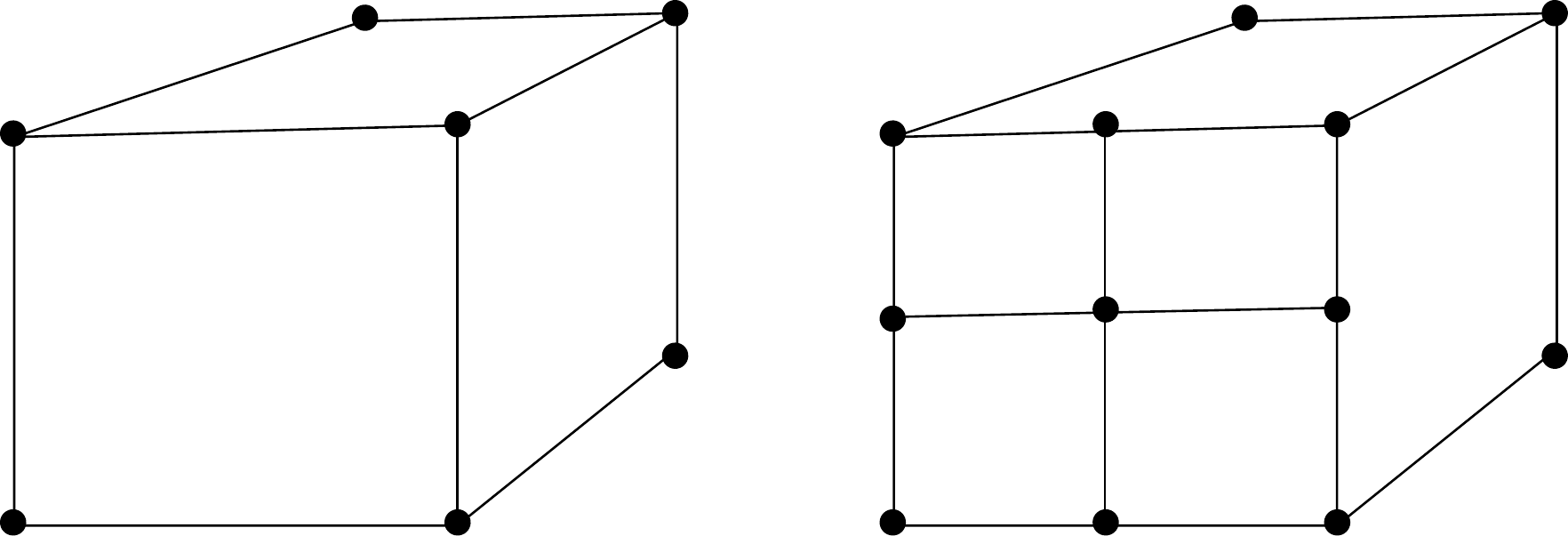}
\end{center}
\caption{The same underlying set $P$ can have different boundary partitions $\tau$, which we count as different polytopes. }\label{Fig:figure_001}
\end{figure}

\noindent We call $\tau$ the \emph{boundary partition} of the polytope, and it supplies us with a way in which to organise the boundary $\partial P$ into 3- (and lower) faces. In particular, the two examples in image \ref{Fig:figure_001} are two different polytopes with the same underlying set $P$.

\section{Subdividing polytopes and sub-polytopes}\label{Sec:SubdividingPolytopes}

In the following we consider moves that coarse grain or refine the boundary partition of a given polytope, i.e.~which changes $\tau$ and $\Pi$, but not $P$ or $\Sigma$. Intuitively, this will be achieved by cutting 3-faces, 2-faces, and 1-faces into smaller pieces. Technically, this is corresponds to redefining which boundary simplices of $\partial\Sigma$ belong to the same faces.

\subsection{Refinement and coarse graining of the boundary partition}

We do regard polytopes with coinciding $P$ but different $\tau$ as different polytopes. However, different boundary partitions $\tau$ can be transformed into one another by coarse graining and/or refining. 

\begin{Definition}
Let $\tau_1:\Pi_1\to \text{Sub}(\Sigma)$ and $\tau_2:\Pi_2\to \text{Sub}(\Sigma)$ be boundary partitions of nice polytopes $(P,\Sigma, \tau_1)$ and $(P,\Sigma,\tau_2)$. Assume there is an order-preserving, level-preserving map $\iota:\Pi_1\to \Pi_2$ such that for a $k$-face $f_1\in\Pi_1$ one has that $\tau_1(f_1)$ is a subcomplex of $\tau_2(\iota(f_1))$. Then $(P,\Sigma, \tau_1)$ is called a \emph{refinement} of  $(P,\Sigma,\tau_2)$.
\end{Definition}

\noindent Intuitively, this means that the boundary faces of the boundary partition $\tau_1$ are the results of subdividing the boundary faces of $\tau_2$. Again, figure \ref{Fig:figure_001} is the prime example for this. 

The notion of refinement induces a partial ordering on the set of boundary partitions of $P$. One can show that there is a smallest element, which is given by the abstract polytope generated by the boundary simplices of $\partial\Sigma$, together with $\tau$ being the identity on $\partial \Sigma$.

For nice polytopes, there is a largest element, which consists of the coarsest boundary possible such that no $3$-faces can be enlarged without violating the condition of being a $3$-ball, or without leaving the common 3d hyperplane. Note that this coarsest boundary partition depends also on $P\subset \mathbb{R}^4$, and is in particular not invariant under p.l.homeomorphisms. 

Due to this, it is enough to consider three moves, which correspond to subdividing 1-, 2- and 3-faces on the boundary. The new boundary partitions $\tau$ are straightforward to construct.

\begin{figure}[hbt!]
\begin{center}
\includegraphics[scale=0.5]{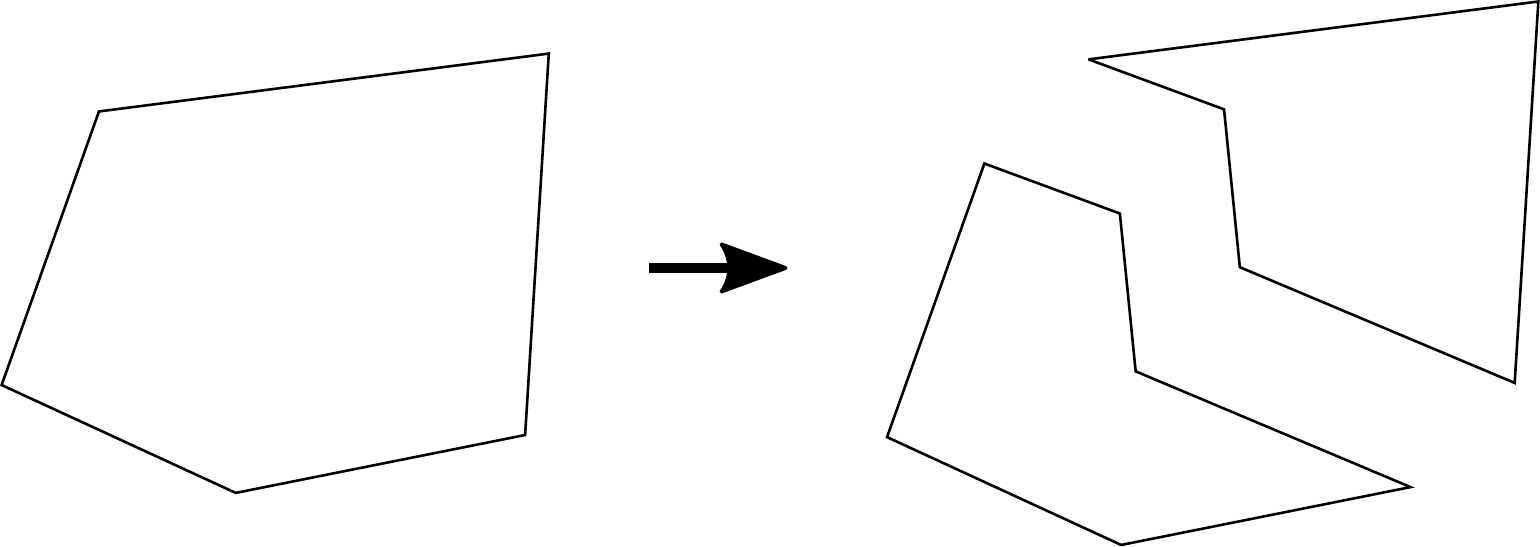}
\end{center}
\caption{Move B: the subdivision of a 2-face into two. Note that this is a change of $\tau$, both $P$ an $\Sigma$ are unchanged. }\label{Fig:figure_SUB_4}
\end{figure}

\begin{figure}[hbt!]
\begin{center}
\includegraphics[scale=0.5]{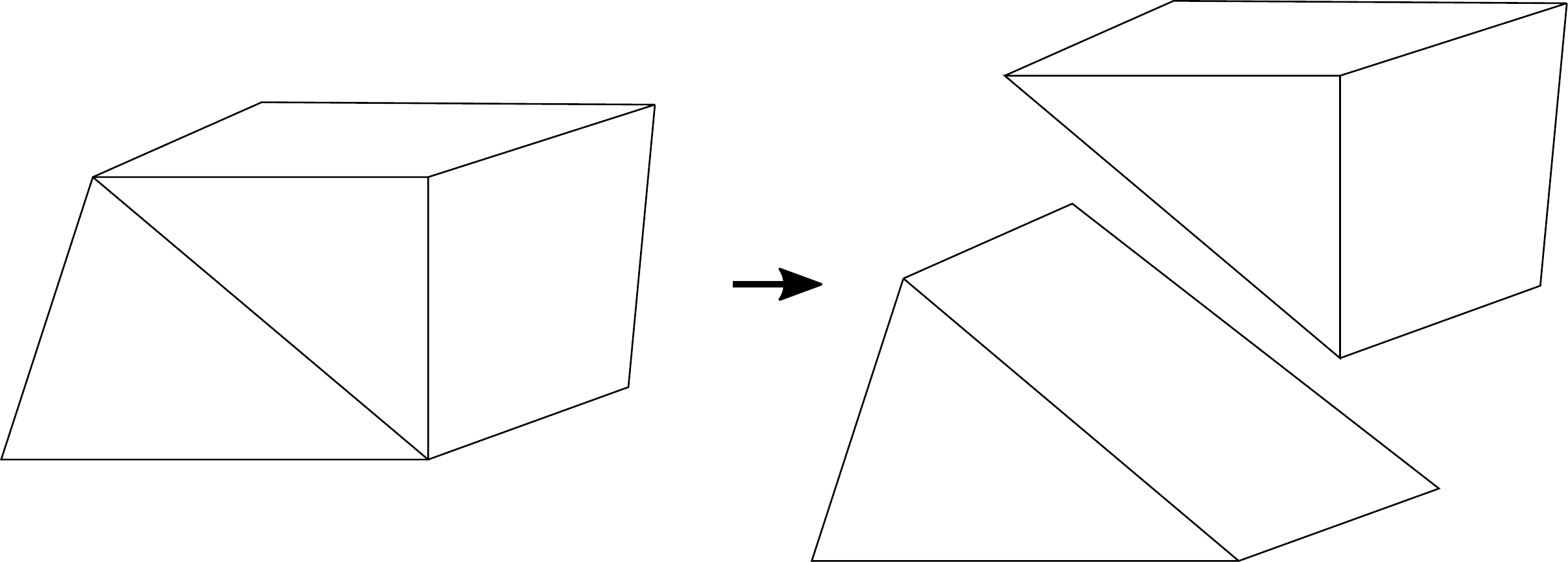}
\end{center}
\caption{Move C: Subdivision of a 3-face.  }\label{Fig:figure_SUB_3}
\end{figure}

\begin{itemize}
\item \textbf{Move A: subdivision of a 1-face:} In this case, one introduces a new 0-face, which splits a 1-face into two.
\item \textbf{Move B: subdivision of a 2-face:} In this case, a 2-face is split into two (figure \ref{Fig:figure_SUB_4}). A collection of new 0- and 1-faces is introduced into $\tau$, comprising the new intersection of the new 2-faces. 
\item \textbf{Move C: subdivision of a 3-face:} The case of subdividing a 3-face into two leads to two new 3-faces, which are meeting at a collection of  new 2-faces (figure \ref{Fig:figure_SUB_3}). 
\end{itemize}

\noindent Notably, the moves $A$, $B$, their inverses, and $C$ can always be performed, if the faces are incident and part of the same face of higher dimension. Only $C^{-1}$ might not be always performed (for a nice polytope), since two neighbouring $3$-faces might not be lying in the same $3d$ hyperplane. It is at this point, where the embedding into $\mathbb{R}^4$ plays a role, and this restriction carries important information about the geometry of $P$, as we will see.

\subsection{Subdividing polytopes}

For convex polytopes there is an easy way to subdivide one polytope into two, which is by intersecting them with half-spaces. Although in principle also possible for the (non-convex) polytopes of our definition, we use the simplicial complex $\Sigma$ for this. 
\begin{Definition}
Let $(P,\Sigma,\tau)$ a nice polytope. The nice polytopes $(P_1,\Sigma_1,\tau_1)$ and $(P_2,\Sigma_2,\tau_2)$ are subdivisions of 
$(P,\Sigma,\tau)$ if the following holds:
\begin{itemize}
\item $P=P_1\cup P_2$,
\item Both $\Sigma_1$ and $\Sigma_2$ are subcomplexes of $\Sigma$,
\item $\Sigma_1\cup \Sigma_2=\Sigma$, and $T:=\Sigma_1\cap \Sigma_2$ is a subcomplex of $\partial\Sigma_1$ and $\partial\Sigma_2$, a combinatorial $3$-ball, called the \emph{glueing surface},
\item The boundaries $\partial\Sigma_1$ and $\partial\Sigma_2$ decompose as 
\begin{align*}
\partial \Sigma_1=T\cup F_1,\qquad \partial\Sigma_2=T\cup F_2,
\end{align*}
\noindent where $S:=T\cap F_1=T\cap F_2=F_1\cap F_2$ is a 2-dim subcomplex which is a topological 2-sphere, being the boundary of either topological 3-ball  $F_1$, $F_2$, $T$.
\item For any 3-face $f\in \Pi$, $\tau(f)$ is either completely in $\Sigma_1$ or completely in $\Sigma_2$. Any lower face has a image under $\tau$ which is either completely in $\Sigma_1$, completely in $\Sigma_2$, or in $S$. 
\item There is a 1-1 correspondence of faces of $\tau_1^{-1}(T)\subset \Pi_1$ and $\tau_2^{-1}(T)\subset \Pi_2$, i.e.~they are isomorphic as sub-polytopes.
\end{itemize}
\end{Definition}

\begin{figure}[hbt!]
\begin{center}
\includegraphics[scale=0.5]{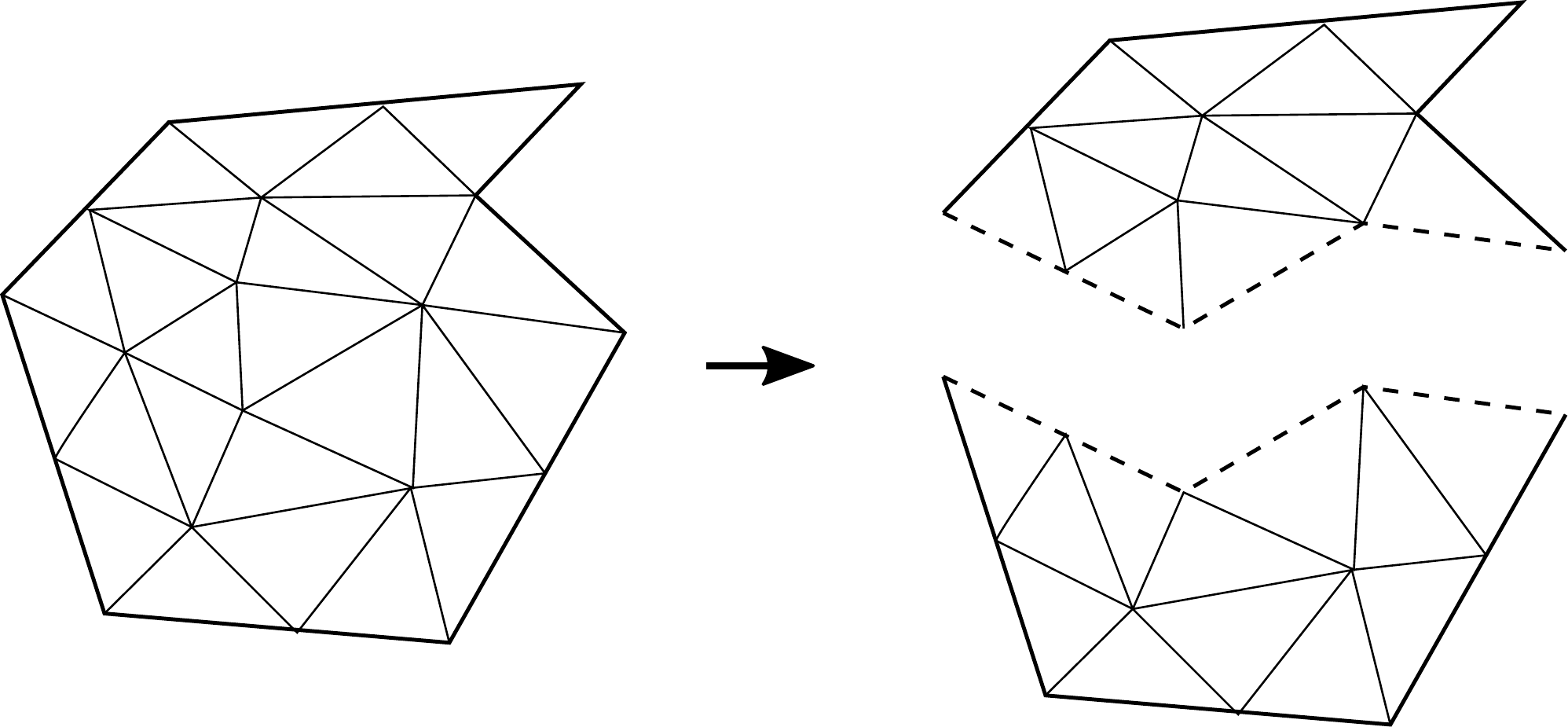}
\end{center}
\caption{Subdividing one polytope into two. The glueing surface, depicted as dashed line, is a sub-complex $T$, which is the triangulation of a 3-ball. }\label{Fig:figure_SUB}
\end{figure}

\noindent Figure \ref{Fig:figure_SUB} depicts the subdivision of a polytope into two. Intuitively, the polytope is cut along $3d$-surfaces which consist of bulk 3-simplices of $\Sigma$, such that they form part of the respective boundaries of $P_1$ and $P_2$. 

\subsection{Constructing polytopes}

There is a smallest polytope within the definition that we have given. It consists of a single 4-simplex $P_\sigma$ embedded in $\mathbb{R}^4$, together with the simplicial complex $\Sigma$ generated by it and its sub-simplices, as well as the polytope generated by $\Sigma$.

Evidently, to construct an arbitrary polytope $(P,\Sigma,\tau)$, one has to  successively glue together the  $P_\sigma$ for all 4-simplices $\sigma\in\Sigma$, ending up with $(P,\Sigma,\tilde{\tau})$, where $\tilde{\tau}$ is the smallest element in the partially ordered set of boundary partitions of $P$. By successively coarse graining the boundary partition, one arrives at $(P,\Sigma,\tau)$.

%This shows that refining or coarse graining the boundary does not change $V$. A posteriori this is clear, since $P$ does not change in any of this, and $V$ will turn out to be the volume of $P$. However, it seems noteworthy, since $\Gamma$ dies change for any of these moves, so a priori it is not at all obvious that (\ref{Eq:VolumeFormula}), which depends in the boundary graph, is invariant. 

\section{Boundary graphs}\label{Sec:BoundaryGraphs}

\noindent For any nice polytope $(P,\Sigma,\tau)$, one can construct a graph $\Gamma\subset S^3$, which is embedded in $S^3$, and is unique up to ambient isotopy of $S^3$. Combinatorially, it is the dual graph to $\Pi$, while the embedding information is determined by the whole polytope. 

\begin{Definition}
Let $(P,\Sigma, \tau)$ be a nice polytope. Then $\partial P\simeq S^3$, and we construct a graph $\Gamma$ in $\partial P$ the following way:
\begin{itemize}
\item For every $3$-face $n\in \Pi$, choose a point in the interior of $\phi(\tau(n))$ as a node of $\Gamma$.
\item Every $2$-face $\ell\in \Pi$ is part of exactly two $3$-faces $n_1, n_2\in \Pi$. Since by construction $\phi(\tau(f))$ is a combinatorial 2-ball which sits in the intersection of the combinatorial $3$-balls $\phi(\tau(n_1))$ and $\phi(\tau(n_2))$, we can choose a path in its interior, and connect the node for $n_1$ and $n_2$ by straight lines passing through that point. 
\end{itemize}
\noindent The resulting graph $\Gamma\subset \partial P$ is defined up to different choices of interior points of combinatorial $k$-balls, from which one can see that $\Gamma$ is uniquely defined up to ambient isotopy in $S^3$.
\end{Definition}

\begin{figure}[hbt!]
\begin{center}
\includegraphics[scale=0.5]{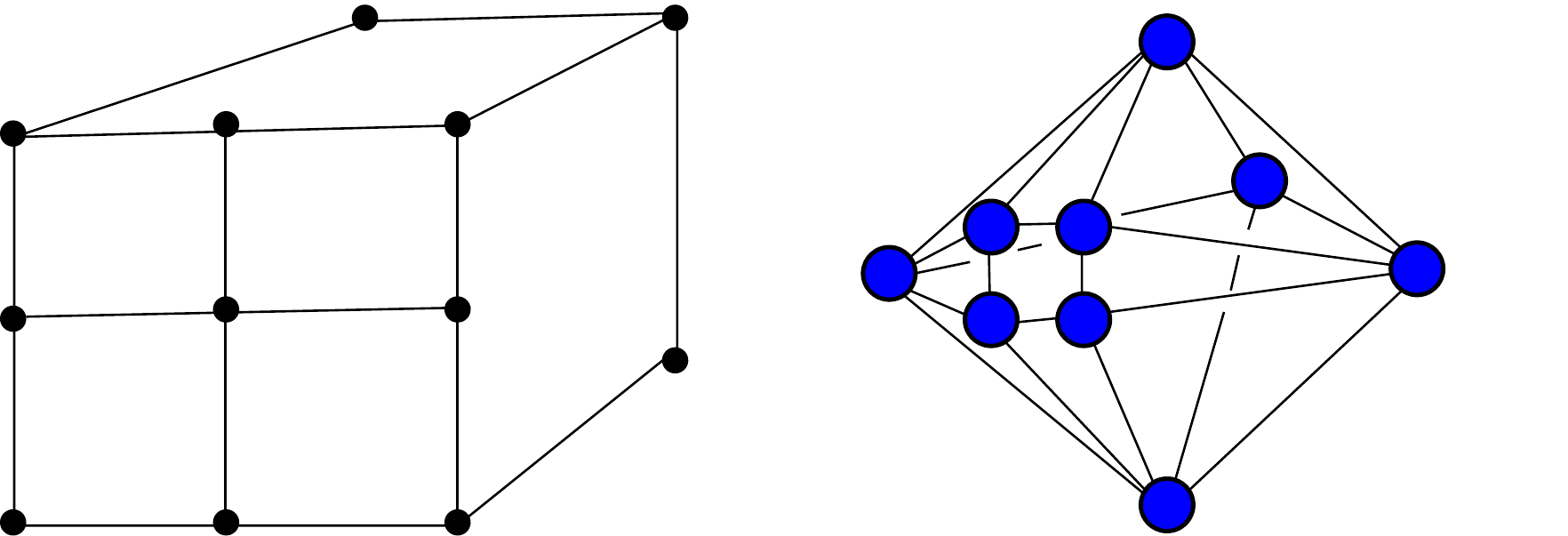}
\end{center}
\caption{A polytope and its boundary graph. The image is a $3d$ representation, which is why the boundary graph can be embedded in $S^2$ rather than $S^3$. }\label{Fig:figure_002}
\end{figure}

\noindent The boundary graph does not only contain the combinatorial information of $\tau$, i.e.~the 2- and 3-faces of the abstract polytope $\Pi$, but also, via its embedding into $S^3$, information about the geometry of the set $P\subset \mathbb{R}^4$. It is this information, particularly the knottings of the graph, which will allow to compute the volume of $P$ from colourings of $\Gamma$. 

\subsection{Bivector geometries}

We define a bivector geometry as an association of 4d bivectors $B_\ell\in\mathbb{R}^4\wedge \mathbb{R}^4$ to the oriented link\footnote{We understand that the reversal of any orientation of a link $\ell$ results in $B_\ell\to -B_{\ell}$.} $\ell$ of an  embedded graph $\Gamma\subset S^3$, which satisfies the following properties:
\begin{itemize}
\item Diagonal simplicity: For each link $\ell$ one has $B_\ell\wedge B_\ell=0$.
\item Cross simplicity: For each pair of links $\ell,\ell'$ meeting at the same node $n$ one has $B_\ell\wedge B_{\ell'}=0$.
\item Closure: For each node $n$ one has 
\begin{align}\label{Eq:ClosureConstraint}
\sum_{\ell\supset n}[n,\ell]\,B_\ell\;=\;0,
\end{align}
\noindent where $[n,\ell]=\pm 1$ if $\ell$ is outgoing from/incoming to the node $n$.
\end{itemize}

\begin{figure}[hbt!]
\begin{center}
\includegraphics[scale=1.0]{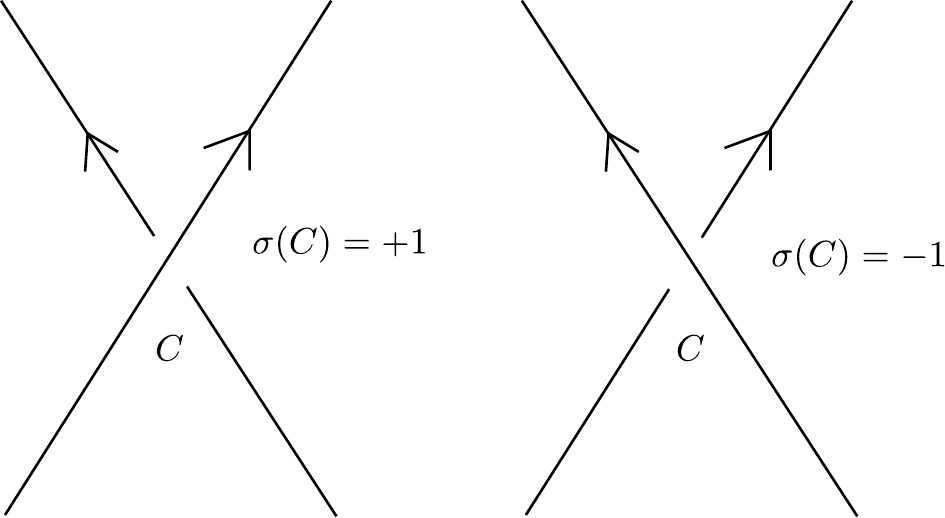}
\end{center}
\caption{There are two types of crossings $C$ within oriented graphs, which are weighted with opposite signs $\sigma(C)=\pm 1$.}\label{Fig:figure_00}
\end{figure}

\noindent Each such graph can be projected from $S^3$ onto the 2d plane, to yield a graph with crossings $C$. For an oriented graph, there are two different types of crossings (see figure \ref{Fig:figure_00}), which have either positive or negative sign $\sigma(C)=\pm 1$. For such a projection, one can define the real number
\begin{align}\label{Eq:VolumeFormula}
V\;=\;\frac{1}{6}\sum_{C}\sigma(C)*\big(B_{\ell_1}\wedge B_{\ell_2}\big),
\end{align}

\noindent where $*$ denotes the Hodge dual of $\wedge^4\mathbb{R}^4$ to $\mathbb{R}$, and $\ell_{1,2}$ are the two links participating in the crossing. In \cite{Bahr:2018pfj} it was shown that (\ref{Eq:VolumeFormula}) is invariant under Reidemeister moves of the graph,  so that $V$ is an invariant of $\Gamma$ itself, independent of the precise projection. 

Every nice polytope induces a bivector geometry on its boundary graph the following way: The links $\ell$ in $\Gamma$ are in one-to-one correspondence with $2$-faces $f$ in $\Pi$. In turn, $\tau(f)$ is a collection of 2-simplices which form part of the boundary of both $\tau(n_1)$ and $\tau(n_2)$, where $\ell$ is a link between the nodes $n_1$ and $n_2$. Assume we chose an orientation so that $s(\ell)=n_1$ and $t(\ell)=n_2$. Then choose an orientation on $\phi(\tau(f))$ such that it is positive as seen from $\phi(\tau(n_1))$. This induces an orientation on each 2-simplex $\phi(\sigma) \subset \phi(\tau(f))$. To each of these oriented $\sigma$, there is a bivector $B_\sigma$ associated to the $2d$ oriented area in $\mathbb{R}^4$ given by $\phi(\sigma)$. 

We then define
\begin{align}\label{Eq:DefinitionBivectorGeometry}
B_\ell\;:=\;\sum_{\sigma\subset\tau(f_\ell)}B_\sigma,
\end{align}

\noindent where the sum ranges over all $2$-simplices in the complex $\tau(f_\ell)$. It can be readily checked that the thus defined bivectors form a bivector geometry on $\Gamma$. For this the conditions that $(P,\Sigma,\tau)$ is a nice polytope is important, in order to ensure the closure condition. 

As a result, for any nice polytope, one can compute the number $V$ as given by (\ref{Eq:VolumeFormula}), which is well-defined by the data $(P,\Sigma,\tau)$. Just as in the case of convex polytopes \cite{Bahr:2018pfj}, we will prove that $V$ is indeed the volume of $P$. First however, we need to check how $V$ changes under change of $\tau$. 

\subsection{Behaviour of $\Gamma$, $B_\ell$, and $V$}

First we note that the refinement or coarse graining of the boundary partition $\tau$ of a polytope $(P,\Sigma,\tau)$ can change $\Gamma$. 
\begin{itemize}
\item \textbf{Move A:} This does not have an effect on $\Gamma$ by definition, since $\Gamma$ only carries information about the 2- and 3-faces of $\Pi$.
\item \textbf{Move B:} This move, and its inverse, do change $\Gamma$ into a new graph $\tilde{\Gamma}$, as one can readily see. The subdivided 2-face corresponds to a link $\ell$ in $\Gamma$, which gets replaced by two links $\ell_1$, $\ell_2$, with the same source and target nodes as $\ell$. These two are, by construction, equivalent to one another  -- and to $\ell$ --  under homotopies which leave the rest of $\Gamma$ unchanged. Since the new 2-faces $f_1$ and $f_2$ comprise the old 2-face $f$, one has, due to (\ref{Eq:DefinitionBivectorGeometry}), that
\begin{align*}
B_\ell\;=\;B_{\ell_1}\,+\,B_{\ell_2}.
\end{align*}

\begin{figure}[hbt!]
\begin{center}
\includegraphics[scale=1.0]{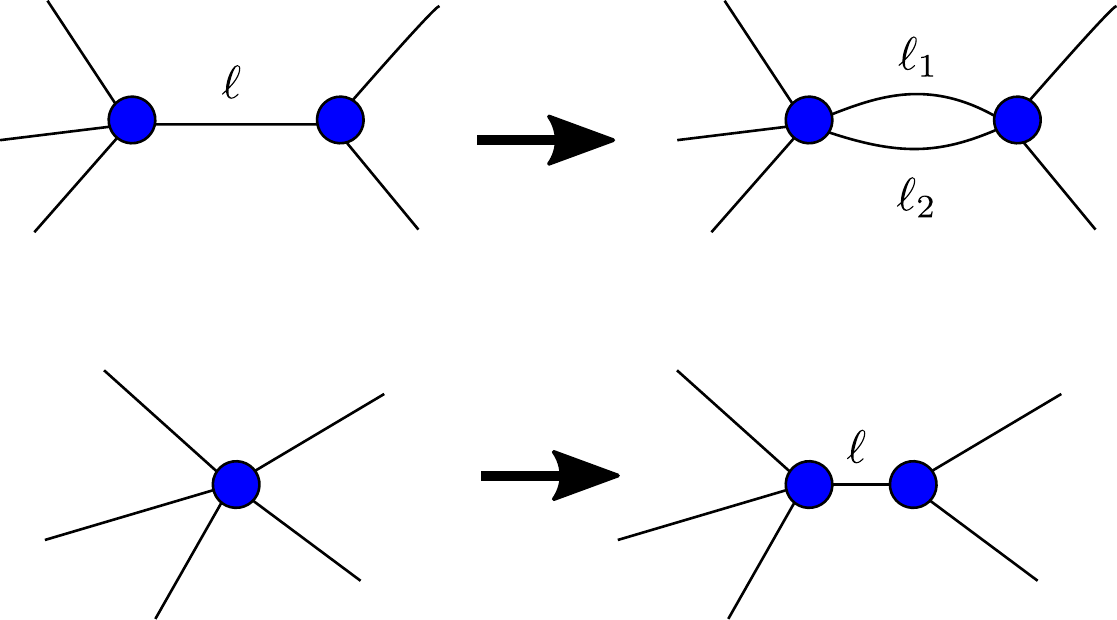}
\end{center}
\caption{Result of move $B$ and $C$ on the boundary graph: For move $B$, the new bivectors satisfy $B_\ell=B_{\ell_1}+B_{\ell_2}$.   }\label{Fig:figure_SUB_5}
\end{figure}

\noindent Since one can rearrange the links of $\Gamma$ in a way that $\ell$ does not partake in any crossing, the same is true for $\ell_1$ and $\ell_2$. So in (\ref{Eq:VolumeFormula}) only crossings between the other links occur, which do not change by the move. Hence $V$ does not change under $B$ or its inverse. 
\item \textbf{Move C:} This move also changes $\Gamma$ into a new graph $\hat{\Gamma}$, where one node gets replaced by two nodes $n_1$ and $n_2$, which are connected by a new link $\ell$. A projection of $\Gamma$ changes into one of $\hat{\Gamma}$ where $\ell$ does not partake in any crossing, while all other crossings do not change, so also for this move, the value of $V$ does not change.
If should be noted that the new bivector geometry is such that even for links $\ell'$,$\ell''$ which meet at $n_1$, $n_2$, respectively, one has that $B_{\ell'}\wedge B_{\ell''}=0$. This shows that the inverse move is only possible if this condition is satisfied for the bivectors of links incident to neighbouring nodes, which need not be the case for arbitrary bivector geometries. This is the analytic analogue of the geometric obstruction to applying $C^{-1}$ discussed in the last section.
\item \textbf{Subdividing polytopes:} If $(P, \Sigma,\tau)$ is subdivided into two $(P_1, \Sigma_1,\tau_1)$, $(P_2, \Sigma_2,\tau_2)$, the new boundary graphs $\Gamma_1$ and $\Gamma_2$ of the two polytopes have an easy relation to one another. In particular, $S$ cuts $\Gamma$ into two pieces (see figure \ref{Fig:figure_SUB_6})
\begin{align*}
\Gamma\;=\;\Gamma_A\;\#\;\Gamma_B,
\end{align*}
\noindent as well as 
\begin{align*}
\Gamma_1\;=\;\Gamma_A\;\#\;\Gamma_T,\qquad \Gamma_2\;=\;\Gamma_B\;\#\;\Gamma_T^*,
\end{align*}

\noindent where $\#$ denotes glueing of graphs along their open ends. The two graphs $\Gamma_T$ and $\Gamma_T^*$ are mirror-images from one another, since they correspond to the part of the boundary which is $T$, however projected to $S^3$ from two different directions in $\mathbb{R}^4$.

\begin{figure}[hbt!]
\begin{center}
\includegraphics[scale=0.5]{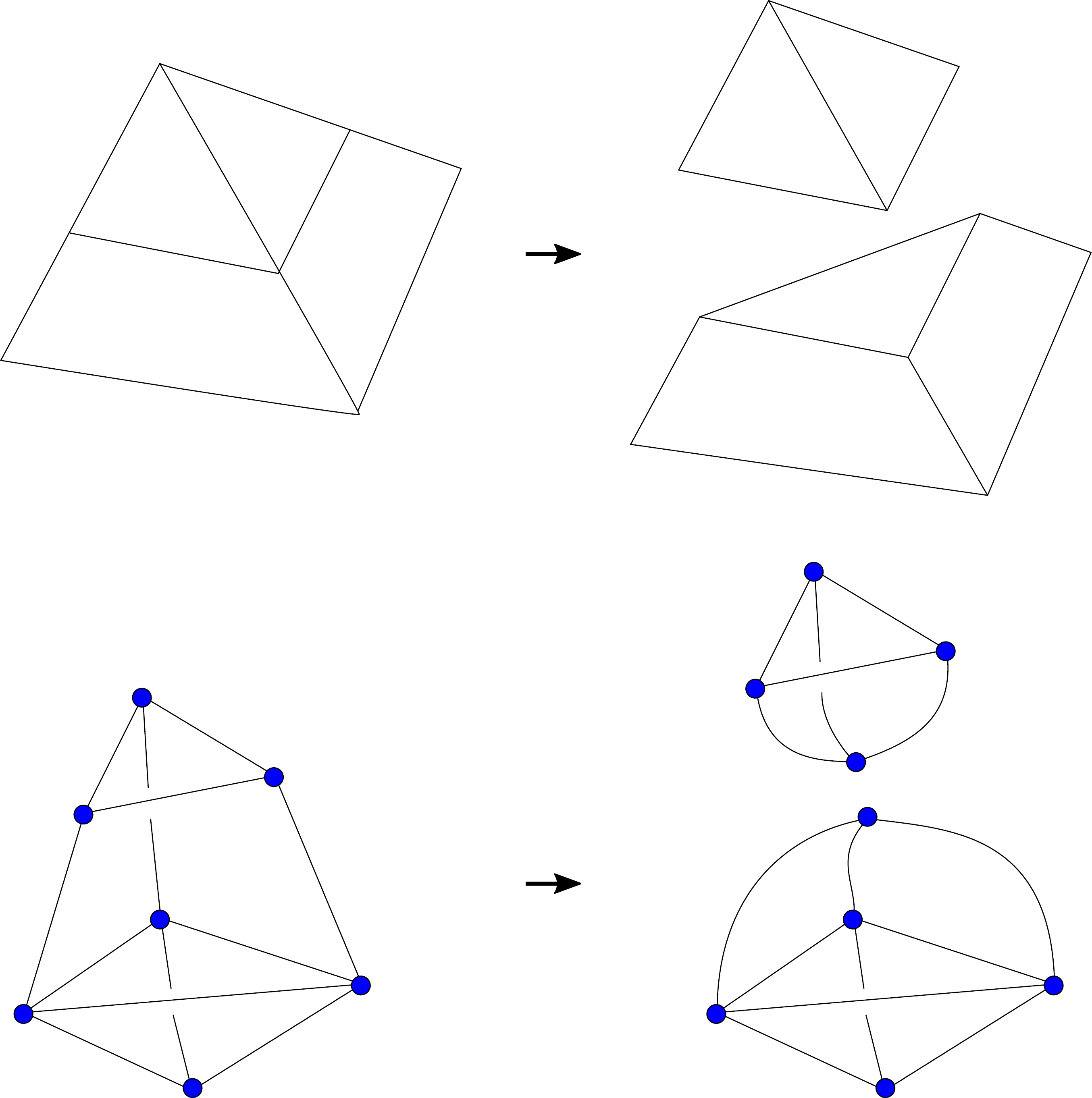}
\end{center}
\caption{The subdivision of a polytope results in a split of the boundary graph $\Gamma$ into two $\Gamma_1$ and $\Gamma_2$.  }\label{Fig:figure_SUB_6}
\end{figure}

It is evident that one can find a projection in which no crossings take place between links of $\Gamma_A$ and $\Gamma_B$.\footnote{This is because $S$ is a 2-sphere which subdivides $S^3$ into two $3$-balls. Wlog $S$ is the equator separating the upper and lower hemisphere, then use stereographic projection w.r.t.~the north- or south pole. }. This means that the crossings in $\Gamma$ are only either in $\Gamma_A$ or $\Gamma_B$, which means that
\begin{align*}
V\;=\;\frac{1}{6}\sum_{C\in \Gamma}V_C\;=\;\frac{1}{6}\sum_{C\in\Gamma_A}V_C\,+\,\frac{1}{6}\sum_{C\in\Gamma_B}V_C
\end{align*}
\noindent where $V_C=\sigma(C)*(B_{1}\wedge B_{2})$. Similarly one has that
\begin{align*}
V_1\;=\;\frac{1}{6}\sum_{C\in\Gamma_A}V_C\,+\,\frac{1}{6}\sum_{C\in\Gamma_T}V_C,\qquad 
V_2\;=\;\frac{1}{6}\sum_{C\in\Gamma_A}V_C\,+\,\frac{1}{6}\sum_{C\in\Gamma_T^*}V_C
\end{align*}
\noindent However, since $\Gamma_T$ and $\Gamma_T^*$ are mirror images of one another, they have identical crossings, just with reversed type (as in figure \ref{Fig:figure_00}). Hence
\begin{align*}
\frac{1}{6}\sum_{C\in\Gamma_T}V_C\;=\;-\frac{1}{6}\sum_{C\in\Gamma_T^*}V_C,
\end{align*}
\noindent which means that
\begin{align*}
V\;=\;V_1\,+\,V_2
\end{align*}
\end{itemize}

\noindent This shows a crucial property of the number $V$: It is invariant under refining or coarse graining the boundary, and it is additive under glueing of polytopes. We have already remarked that every polytope $(P,\Sigma,\tau)$ can be constructed by successively glueing together the $P_\sigma$ for every 4-simplex $\sigma\in\Sigma$, and then coarse graining the boundary partition until one arrives at $\tau$. The number $V(P)$ for the original polytope is therefore the sum of the corresponding numbers $V(P_\sigma)$, which has been shown to equal the 4-volume $Vol(\sigma)$ in \cite{Bahr:2018pfj}. We can conclude:

\begin{Lemma}
Let $(P,\Sigma,\tau)$ be a nice polytope, $\Gamma$ its boundary graph, $\{B_\ell\}_\ell$ the associated bivector geometry, then
\begin{align}\label{Eq:FinalFormula}
Vol(P)\;=\;V\;=\;\frac{1}{6}\sum_{C}\,\sigma(C)*\big(B_1\wedge B_2\big)
\end{align}
\noindent is the 4-volume of $P\subset\mathbb{R}^4$.
\end{Lemma}

\noindent This argument generalises to proof which has been delivered in \cite{Bahr:2018pfj} for convex polytopes to the more general, non-convex ones. 

\section{Non-convex polytopes in spin foam asymptotics}\label{Sec:InSFM}

We come to the second part of this article, which shows how non-convex $4d$ polytopes can arise as semiclassical geometries in the path integral of the EPRL-FK model. The initial setup is a graph $\Gamma\subset S^3$, together with a set of boundary data
\begin{align}\label{Eq:BoundaryDataGeneral}
\big\{j_{ab},\vec{n}_{ab}\big\}_{(ab)},
\end{align}

\noindent where $(ab)$ denote the link from the node $b$ to the node $a$. These labels specify a spin network state $\psi$ in the boundary Hilbert space associated to $\Gamma$ \cite{Bianchi:2010gc, Livine:2007vk}, and the amplitude associated to this state 
\begin{align*}
\mathcal{A}(\psi)\;:=\;\langle\psi\,|\,\psi_0\rangle
\end{align*}

\noindent  with the $BF$ vacuum state $\psi_0$ is one of the fundamental building blocks of the theory. The precise definition of this can be found e.g.~in \cite{Engle:2007wy, Perez:2012wv}.

In the asymptotic limit where $j_{ab}\to \lambda j_{ab}$ and $\lambda\to\infty$, one can evaluate the expression for $\mathcal{A}$ by means of the extended stationary phase approximation. 

The critical and stationary points here correspond to $SU(2)\times SU(2)$-elements $(g_a^{(+)},g_a^{(-)})$ associated to the nodes of $\Gamma$, and together with the boundary data (\ref{Eq:BoundaryDataGeneral}) these determine a bivector geometry on $\Gamma$ via
\begin{align*}
B_{ab}\;\sim\;j_{ab}
\Big(
g_a^{(+)}\vec{n}_{ab},\,g_a^{(-)}\vec{n}_{ab}
\Big),
\end{align*}

\noindent using the action of $SU(2)$ on $\mathbb{R}^3$ and the isomorphism $\mathbb{R}^4\wedge\mathbb{R}^4\sim \mathbb{R}^3\oplus\mathbb{R}^3$.

In the following, we will give an example of boundary data which results in several bivector geometries, some of them corresponding to non-convex polytopes in the asymptotic limit. One can show this explicitly, but we will use the formula (\ref{Eq:FinalFormula}) to demonstrate the geometry of the asymptotic geometries which occur.

\subsection{Boundary data of a hypercubecube}

To give an example for a set of boundary data which leads to several different asymptotic geometries, we start from the hypercube, and subdivide one of its 3-faces. A slight deformation of the resulting data will be our example.

\begin{figure}[hbt!]
\begin{center}
\begin{minipage}[t]{0.45\textwidth}
\includegraphics[scale=0.35]{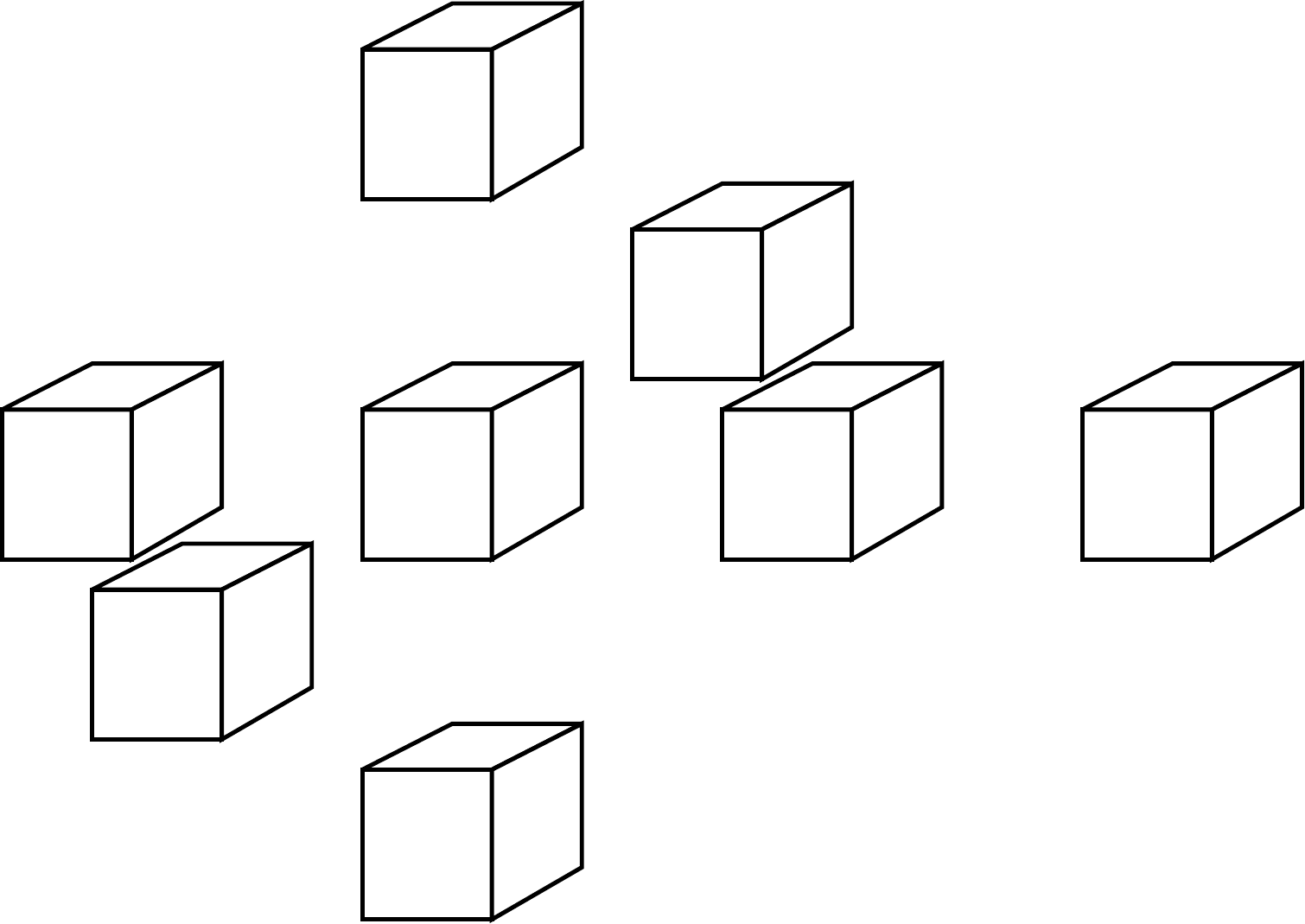}
\end{minipage}
\begin{minipage}[t]{0.45\textwidth}
\includegraphics[scale=0.5]{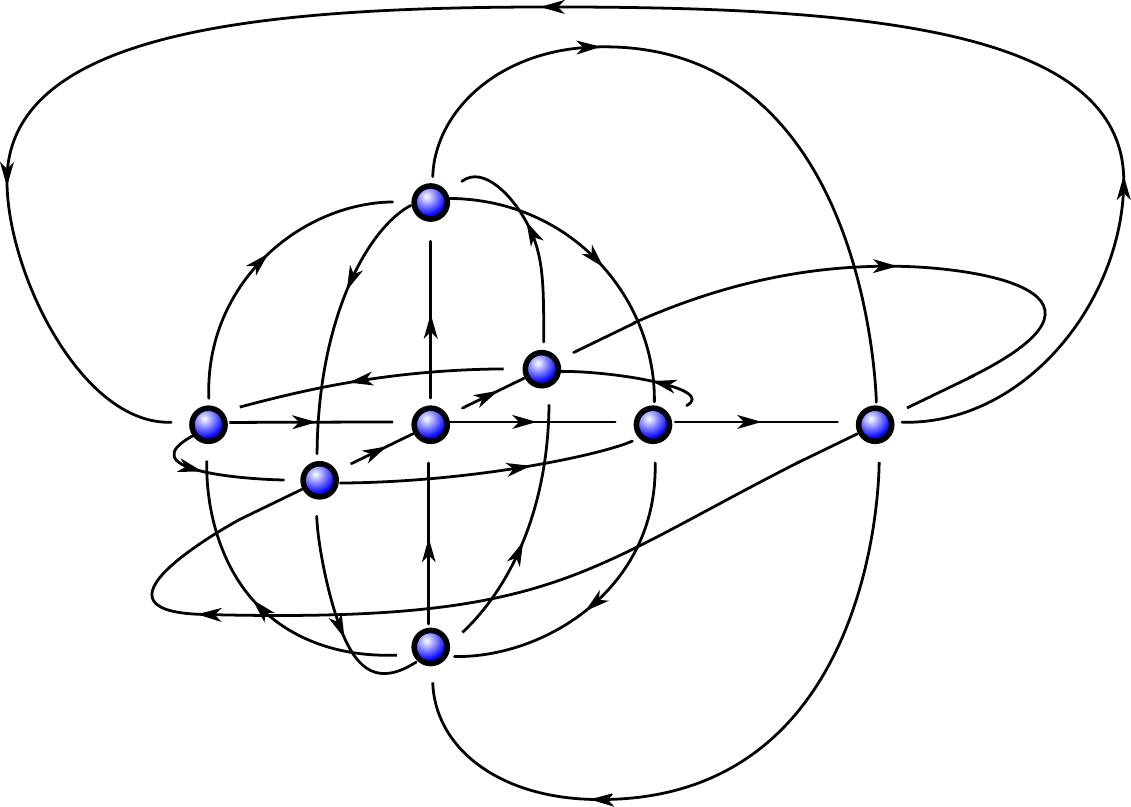}
\end{minipage}
\end{center}
\caption{The boundary of a $4d$ hypercube consists of eight cubes. The associated boundary graph is depicted as well, and in a more organised fashion in figure \ref{Fig:figure_04}.}\label{Fig:figure_03}
\end{figure}

\begin{figure}[hbt!]
\begin{center}
\includegraphics[scale=1.0]{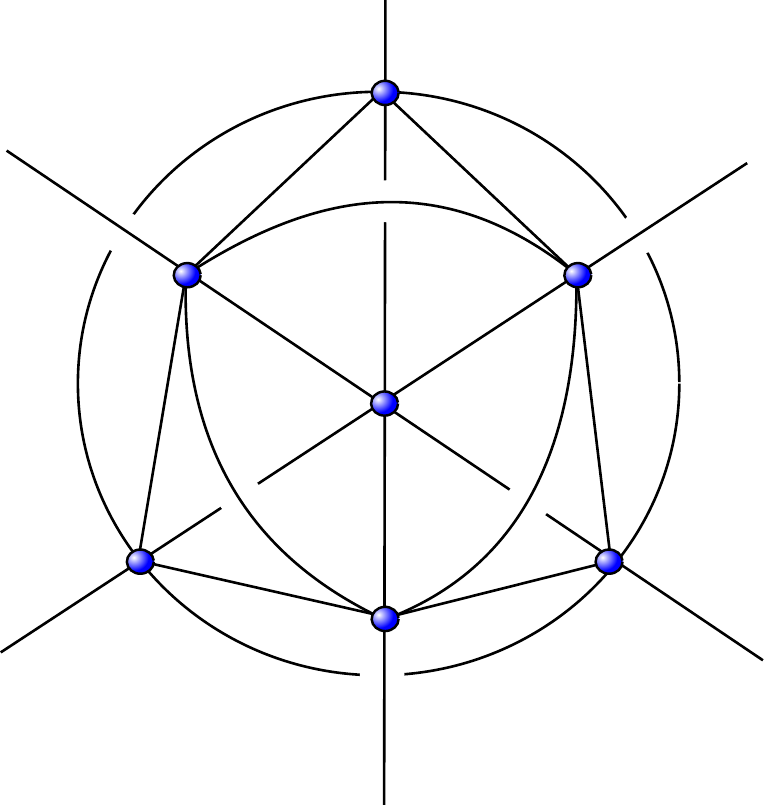}
\end{center}
\caption{The boundary graph of a $4d$ hypercube consists of eight nodes and 24 links. One of the nodes has been placed at infinity. In this form, the graph contains six crossings, all of which have $\sigma(C)=1$, when using the link orientations from figure \ref{Fig:figure_03}.}\label{Fig:figure_04}
\end{figure}

The hypercube has a boundary consisting of eight cubes (figure \ref{Fig:figure_03}). Its boundary graph is depicted in figures \ref{Fig:figure_03} and \ref{Fig:figure_04}.

\noindent For simplicity, we choose all spins equal to $j$, so that each 3d boundary polyhedron is a cube, with $3d$ normals $\pm\vec{e}_i$, $i=1,2,3$. The original rotation of the cubes is depicted in figure \ref{Fig:figure_03}. Normals of opposing faces point in opposite directions. 
\begin{align}\label{Eq:BoundaryDataCube}
j_{ab}\;=\;j,\qquad,\vec{n}_{ab}\;=\;\pm \vec{e}_i,\qquad i=1,2,3
\end{align}

\noindent To understand the geometry of the Euclidean EPRL-FK-KKL model for $\gamma<1$, it is sufficient to investigate the associated $SU(2)$-BF theory spin foam model. The variables are $g_a\in SU(2)$, where $a$ ranges through all nodes of the boundary graph. The equations for critical, stationary geometries in the path integral can then be formulated as

\begin{align}\label{Eq:CritStatEqns}
g_a\vec{n}_{ab}\;=\;-g_b\vec{n}_{ba},
\end{align}

\noindent where $\vec{n}_{ab}$ is the $3d$ normal of the polyhedron at $a$, associated to the $2d$ face where it is connected to the polyhedron $b$ (see figure \ref{Fig:figure_03}). Every solution has a global symmetry $g_a\to h\,g_a$ for all $a$, which is why one usually fixes one $g_a=\mathds{1}$. We do this for $a=0$, which corresponds to the ``central'' node in figure \ref{Fig:figure_04}. Due to the action of $SU(2)$ on vectors $\vec{n}\in\mathbb{R}^3$, $(-g)\vec{n}=g\vec{n}$. One can therefore denote the solutions modulo signs.\footnote{Indeed, the equations (\ref{Eq:CritStatEqns}) are actually equations in $SO(3)$.}

The solutions consist of rotations around the main axes by $\frac{\pi}{2}$ for $a=1,\ldots,6$, and a rotation of $\pi$ around the 1-axis for $a=7$, which in figure \ref{Fig:figure_04} corresponds to the node at infinity. The results are summarised in table \ref{Tab:Table01}.
\begin{figure}
\begin{center}
\begin{tabular}{c|c|c}
& Solution $+$ & Solution $-$  \\[5pt] \hline
$g_1$ & $\exp\left(i\frac{\pi}{4}\sigma_1\right)$ & $\exp\left(-i\frac{\pi}{4}\sigma_1\right)$ \\[5pt]
$g_2$ & $\exp\left(i\frac{\pi}{4}\sigma_2\right)$ & $\exp\left(-i\frac{\pi}{4}\sigma_2\right)$ \\[5pt]
$g_3$ & $\exp\left(i\frac{\pi}{4}\sigma_3\right)$ & $\exp\left(-i\frac{\pi}{4}\sigma_3\right)$ \\[5pt]
$g_4$ & $\exp\left(-i\frac{\pi}{4}\sigma_1\right)$ & $\exp\left(i\frac{\pi}{4}\sigma_1\right)$ \\[5pt]
$g_5$ & $\exp\left(-i\frac{\pi}{4}\sigma_2\right)$ & $\exp\left(i\frac{\pi}{4}\sigma_2\right)$ \\[5pt]
$g_6$ & $\exp\left(-i\frac{\pi}{4}\sigma_3\right)$ & $\exp\left(i\frac{\pi}{4}\sigma_3\right)$ \\[5pt]
$g_7$ & $\exp\left(i\frac{\pi}{2}\sigma_1\right)$ & $\exp\left(-i\frac{\pi}{2}\sigma_1\right)$
\end{tabular}
\caption{Critical stationary points of the boundary data (\ref{Eq:BoundaryDataCube}). We remind the reader that the action of an $SU(2)$-element $\exp(-i\theta \vec{n})$ on $\mathbb{R}^3$ is a rotation around the axis $\vec{n}$ with angle $2\theta$.}\label{Tab:Table01}
\end{center}
\end{figure}

\noindent As is usual in this case, there are two distinct solutions $g_a^{(\pm)}$, which can be combined to a 4d geometry in terms of bivectors $B_{ab}\sim j(\,g_a^{(+)}\vec{n}_{ab},\,\,g_a^{(-)}\vec{n}_{ab})$.

\subsection{Boundary data of a $4d$ ``House of Santa Claus''}

\begin{figure}[hbt!]
\begin{center}
\includegraphics[scale=0.5]{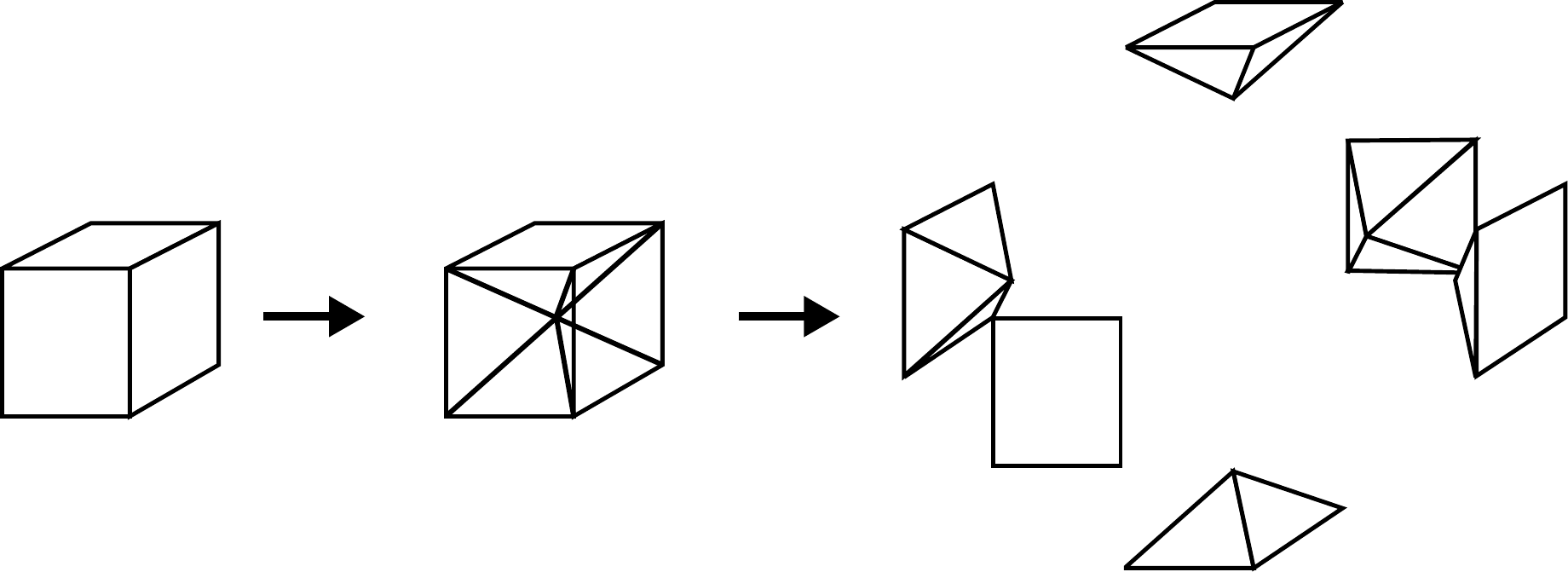}
\end{center}
\caption{Subdividing one cube into six pyramids changes the boundary graph.}\label{Fig:figure_05}
\end{figure}

\begin{figure}[hbt!]
\begin{center}
\includegraphics[scale=1.0]{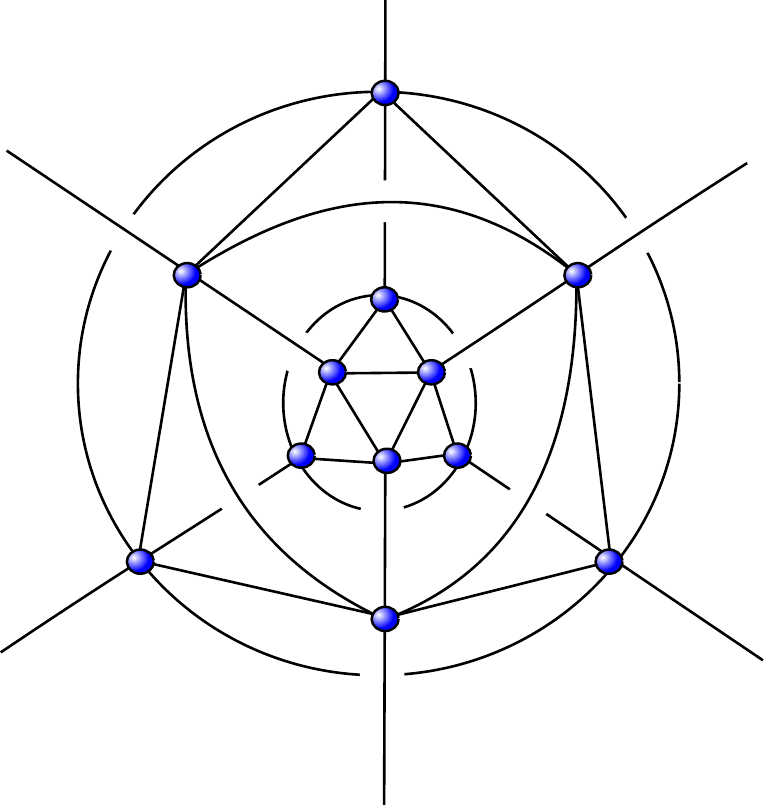}
\end{center}
\caption{Subdividing the central cube changes the boundary graph from figure \ref{Fig:figure_04} to the one depicted here. It gains three more crossings, all of which have $\sigma(C)=+1$.}\label{Fig:figure_06a}
\end{figure}

To produce a set of boundary data which admits more than one $4d$ geometry, we subdivide the cube of the $0$-node into six pyramids, as in figure \ref{Fig:figure_05}. The new boundary graph is depicted in figure \ref{Fig:figure_06a}. 

\begin{figure}[hbt!]
\begin{center}
\includegraphics[scale=0.5]{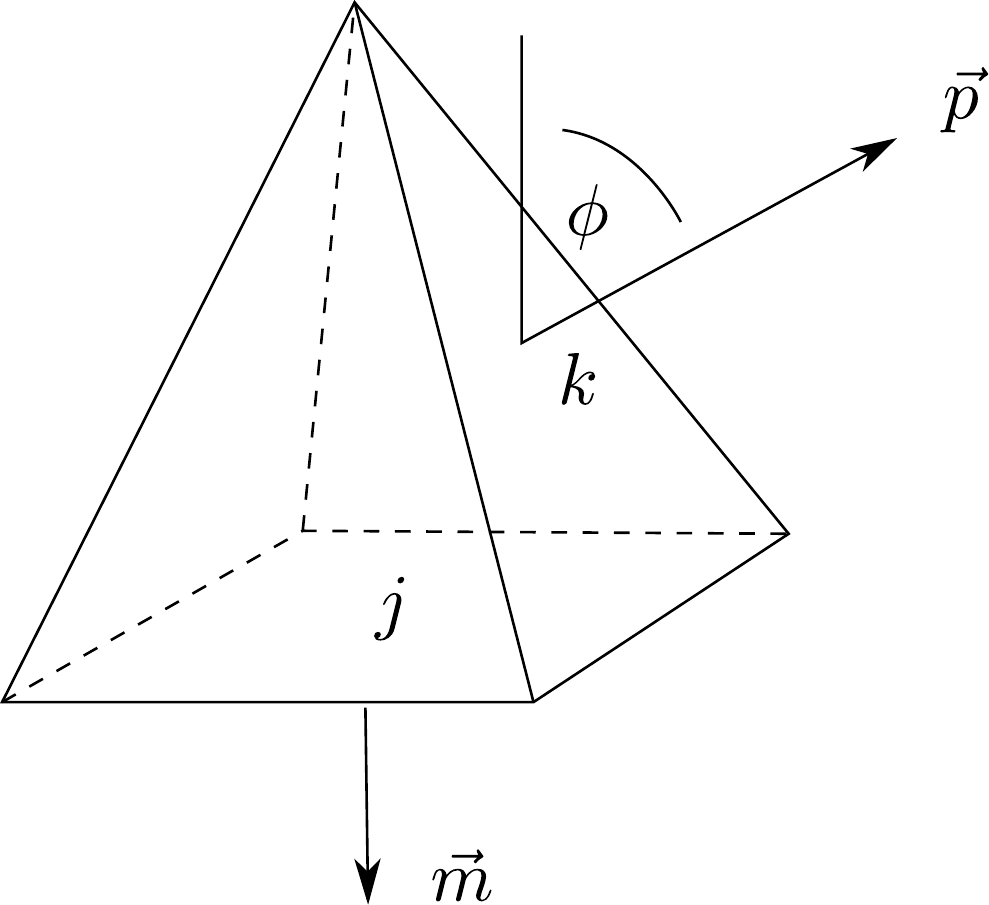}
\end{center}
\caption{One pyramid of the new boundary data. The angle $\phi$ is determined by $j$ and $k$. }\label{Fig:figure_XY}
\end{figure}

\noindent The links in the graph depicted in figure \ref{Fig:figure_06a} which belong to the triangular faces get assigned the spin $k$. The normals can be read off in terms of $j$ and $k$ (figure \ref{Fig:figure_XY}), and can be framed in terms of the angle $\phi$, satisfying
\begin{align*}
\cos\phi\;=\;\frac{k}{4j},
\end{align*}

\noindent which follows from the fact that the data satisfy the closure constraint (\ref{Eq:ClosureConstraint}). The data therefore can be completely framed in terms of the angle $\phi$ and the spin $j$, or, equivalently, in terms of the two spins $j$ and $k$.

We numerate cubes by $a=1,\ldots, 7$, pyramids by $n=1,\ldots 6$. The new data consist of normals $\vec{n}_{ab}$ between cubes, normals $\vec{m}_{an}$ between cubes and pyramids, and normals $\vec{p}_{nm}$ between pyramids (whenever the graph allows this -- cube $7$ has no connection to any pyramid, for instance). The data
\begin{align}\label{Eq:NewData}
j,k,\vec{n}_{ab},\,\vec{m}_{na},\,\vec{p}_{nm},\qquad a=1,\ldots, 7,\,n=1,\ldots, 6
\end{align}

\noindent depends on $\phi$, and corresponds to the original geometry if $\phi=\frac{\pi}{4}$. However, one can a priori choose any other angle between $0$ and $\frac{\pi}{2}$. This would correspond to more acute or obtuse angles in figure \ref{Fig:figure_XY}.

We will now consider the asymptotic geometries depending on $\phi$. We remove $g_0$ from the list of variables, and replace it with six new variables $h_{n}$, $n=1,\ldots, 6$. They correspond to the six pyramids, i.e.~the central nodes in figure \ref{Fig:figure_06a}. 

The critical and stationary equations for this new set of boundary data is still given by 
\begin{align}\label{Eq:CritStatEqns_OuterOuter}
g_a\vec{n}_{ab}\;&=\;-g_b\vec{n}_{ba},\\[5pt]\label{Eq:CritStatEqns_OuterInner}
g_a\vec{m}_{an}\;&=\;-h_{n}\vec{m}_{na},\\[5pt]\label{Eq:CritStatEqns_InnerInner}
h_{n}\vec{p}_{nm}\;&=\;-h_{m}\vec{p}_{mn}
\end{align}

\noindent for $a,b=1,\ldots, 7$ and $n,m=1\ldots, 6$, whenever the appropriate links exist in the graph. One can see that the original gauge fixed equations (\ref{Eq:CritStatEqns}) are a subset (\ref{Eq:CritStatEqns_OuterOuter}) of the new ones, so that the solutions for $g_a$ in table \ref{Tab:Table01} also solve the new equations. Also, $h_{1,6}$ have to be  rotations around the 1-axis, $h_{2,5}$ around the 2-axis, and $h_{3,4}$ around the 3-axis, which solves the equations (\ref{Eq:CritStatEqns_OuterInner}) between the $g_a$ and the $h_n$. The equations (\ref{Eq:CritStatEqns_InnerInner}) remain, and as an example, we consider the equations $n=1$ and $m=2$. We can read off the normals from image (\ref{Fig:figure_XY}) as
\begin{align}
\vec{p}_{12}\;=\;
\begin{pmatrix}
-\cos\phi \\ \sin\phi \\ 0 
\end{pmatrix},\qquad
\vec{p}_{21}\;=\;
\begin{pmatrix}
\sin\phi \\ -\cos\phi \\ 0 
\end{pmatrix}.
\end{align}

\noindent Since the rotation axes of $h_1=\exp(-i\theta_1 \sigma_1/2)$ and $h_2=\exp(-i\theta_2\sigma_2/2)$ are fixed by (\ref{Eq:CritStatEqns_OuterInner}), we obtain the equations
\begin{align*}
-\cos\phi\;&=\;-\sin\phi\cos\theta_2\\[5pt]
\sin\phi\cos\theta_2\;&=\;\cos\phi\\[5pt]
\sin\phi\sin\theta_1\;&=\;\sin\phi\sin\theta_2.
\end{align*}

\noindent These are only two independent equations, since the sum of their norm squares is always equal to $1$. Since the angles $\phi$ are restricted to lie within $\phi\in(0,\pi/2)$, $\tan\phi>0$, so both $\theta_1$ and $\theta_2$ have to lie within the interval $(-\pi/2,\pi/2)$. With this information we can read off
\begin{align}\label{Eq:ThetaInTermsOfPhi}
\theta_1\;=\;\theta_2\;\equiv\theta,\qquad \tan\phi\,\cos(\theta)\;=\;1.
\end{align}

\noindent Due to symmetry, one can show that the solution for the other $h_m$ is similar, where the rotation angles for all $h_m$ equal  $\theta$. 

In order for $\theta$ to exist and be real, the angle $\phi$ has to be restricted to lie within $\phi\in[\pi/4,\pi/2)$. So one can deform the pyramids by making the angle at the tip more acute. By making it more obtuse, one prescribes boundary data for which there are no solutions to the critical and stationary equations (\ref{Eq:CritStatEqns_OuterOuter}) -- (\ref{Eq:CritStatEqns_InnerInner}), so the Euclidean amplitude would be exponentially suppressed in the large spin limit.\footnote{It should be noted that this sort of boundary data would prescribe a Lorentzian pyramid, and might serve as boundary data for the Lorentzian amplitude.} 

It should be noted that $\theta$ is only determined up to a sign, so there are two solutions. It is also noteworthy that this sign can be chosen \emph{independently} of the $(+/-)$-solution for the $g_a$. So the choice of the $(+/-)$-solution for the $g_a$, as well as the sign of $\theta$, allow for four solutions in total, which are summarised in table \ref{Tab:Table02}

\begin{figure}
\begin{center}
\begin{tabular}{c|c|c|c|c}
&  $\Sigma_1$ $\sim(++)$ &  $\Sigma_2$ $\sim(-+)$ &  $\Sigma_3$ $\sim(+-)$ &  $\Sigma_4$ $\sim(--)$\\[5pt] \hline
$g_1$ & $\exp\left(i\frac{\pi}{4}\sigma_1\right)$ & $\exp\left(-i\frac{\pi}{4}\sigma_1\right)$ &  $\exp\left(i\frac{\pi}{4}\sigma_1\right)$ & $\exp\left(-i\frac{\pi}{4}\sigma_1\right)$ \\[5pt]
$g_2$ & $\exp\left(i\frac{\pi}{4}\sigma_2\right)$ & $\exp\left(-i\frac{\pi}{4}\sigma_2\right)$ &  $\exp\left(i\frac{\pi}{4}\sigma_2\right)$ & $\exp\left(-i\frac{\pi}{4}\sigma_2\right)$\\[5pt]
$g_3$ & $\exp\left(i\frac{\pi}{4}\sigma_3\right)$ & $\exp\left(-i\frac{\pi}{4}\sigma_3\right)$ &  $\exp\left(i\frac{\pi}{4}\sigma_3\right)$ & $\exp\left(-i\frac{\pi}{4}\sigma_3\right)$ \\[5pt]
$g_4$ & $\exp\left(-i\frac{\pi}{4}\sigma_1\right)$ & $\exp\left(i\frac{\pi}{4}\sigma_1\right)$ &  $\exp\left(-i\frac{\pi}{4}\sigma_1\right)$ & $\exp\left(-i\frac{\pi}{4}\sigma_1\right)$ \\[5pt]
$g_5$ & $\exp\left(-i\frac{\pi}{4}\sigma_2\right)$ & $\exp\left(i\frac{\pi}{4}\sigma_2\right)$ &  $\exp\left(-i\frac{\pi}{4}\sigma_2\right)$ & $\exp\left(i\frac{\pi}{4}\sigma_2\right)$ \\[5pt]
$g_6$ & $\exp\left(-i\frac{\pi}{4}\sigma_3\right)$ & $\exp\left(i\frac{\pi}{4}\sigma_3\right)$ &  $\exp\left(-i\frac{\pi}{4}\sigma_3\right)$ & $\exp\left(i\frac{\pi}{4}\sigma_3\right)$ \\[5pt]
$g_7$ & $\exp\left(i\frac{\pi}{2}\sigma_1\right)$ & $\exp\left(-i\frac{\pi}{2}\sigma_1\right)$ &  $\exp\left(-i\frac{\pi}{4}\sigma_1\right)$ & $\exp\left(i\frac{\pi}{4}\sigma_1\right)$ \\[5pt]
$h_1$ & $\exp\left(i\frac{\theta}{2}\sigma_1\right)$ & $\exp\left(i\frac{\theta}{2}\sigma_1\right)$ & $\exp\left(-i\frac{\theta}{2}\sigma_1\right)$ & $\exp\left(-i\frac{\theta}{2}\sigma_1\right)$ \\[5pt]
$h_2$ & $\exp\left(i\frac{\theta}{2}\sigma_2\right)$ & $\exp\left(i\frac{\theta}{2}\sigma_2\right)$ & $\exp\left(-i\frac{\theta}{2}\sigma_2\right)$ & $\exp\left(-i\frac{\theta}{2}\sigma_2\right)$ \\[5pt]
$h_3$ & $\exp\left(i\frac{\theta}{2}\sigma_3\right)$ & $\exp\left(i\frac{\theta}{2}\sigma_3\right)$ & $\exp\left(-i\frac{\theta}{2}\sigma_3\right)$ & $\exp\left(-i\frac{\theta}{2}\sigma_3\right)$ \\[5pt]
$h_4$ & $\exp\left(-i\frac{\theta}{2}\sigma_3\right)$ & $\exp\left(-i\frac{\theta}{2}\sigma_3\right)$ & $\exp\left(i\frac{\theta}{2}\sigma_3\right)$ & $\exp\left(i\frac{\theta}{2}\sigma_3\right)$ \\[5pt]
$h_5$ & $\exp\left(-i\frac{\theta}{2}\sigma_2\right)$ & $\exp\left(-i\frac{\theta}{2}\sigma_2\right)$ & $\exp\left(i\frac{\theta}{2}\sigma_2\right)$ & $\exp\left(i\frac{\theta}{2}\sigma_2\right)$ \\[5pt]
$h_6$ & $\exp\left(-i\frac{\theta}{2}\sigma_1\right)$ & $\exp\left(-i\frac{\theta}{2}\sigma_1\right)$ & $\exp\left(i\frac{\theta}{2}\sigma_1\right)$ & $\exp\left(i\frac{\theta}{2}\sigma_1\right)$ 
\end{tabular}
\caption{Solutions $\Sigma_1$ -- $\Sigma_4$ to the critical stationary points of $SU(2)$ $BF$ theory for boundary data (\ref{Eq:NewData}). The angle $\theta$ is determined via (\ref{Eq:ThetaInTermsOfPhi}). }\label{Tab:Table02}
\end{center}
\end{figure}

The asymptotics of the EPRL-FK model asymptotics have been investigated in many works, for several different sets of boundary data (see e.g.~\cite{Barrett:2009gg,Bahr:2015gxa, Dona:2017dvf, Bahr:2017eyi}). We consider the case $\gamma<1$, where one can read off the critical and stationary points directly from those of the associated $SU(2)$-BF theory, which we have computed in table \ref{Tab:Table02}. The gauge group underlying the model is $SU(2)\times SU(2)$ for the Euclidean signature, and for $\gamma<1$ the amplitude factorises into two $SU(2)$-BF amplitudes
\begin{align}\label{Eq:AmplitudeFactorisation}
\mathcal{A}_{\text{EPRL-FK}}\;=\;\mathcal{A}^{(+)}_{SU(2)}\;\mathcal{A}^{(-)}_{SU(2)}
\end{align}

\noindent The asymptotics of $\mathcal{A}_\text{EPRL-FK}$ can thus be computed by the product of the asymptotic expression. Both of the $\pm$-amplitudes in (\ref{Eq:AmplitudeFactorisation}) have the same boundary vectors $\vec{n}$, but different boundary spins $j^\pm=|1\pm\gamma|/2$. Since the critical and stationary equations (\ref{Eq:CritStatEqns_OuterOuter} -- \ref{Eq:CritStatEqns_InnerInner}) only depend on the $\vec{n}$, the critical stationary points coincide as elements of $SU(2)$. Since there are four of these points (listed in table \ref{Tab:Table01}), the asymptotic expression for $\mathcal{A}$ is a sum over 16 terms, consisting of pairs of solutions from table \ref{Tab:Table01}.
\begin{align*}
\mathcal{A}\;&\sim\;\left(\mathcal{A}^{(+)}(\Sigma_1)+\mathcal{A}^{(+)}(\Sigma_2)+\mathcal{A}^{(+)}(\Sigma_3)+\mathcal{A}^{(+)}(\Sigma_4)\right)\\[5pt]
&\qquad\times\left(\mathcal{A}^{(-)}(\Sigma_1)+\mathcal{A}^{(-)}(\Sigma_2)+\mathcal{A}^{(-)}(\Sigma_3)+\mathcal{A}^{(-)}(\Sigma_4)\right)
\end{align*}

\noindent The geometric interpretation of each of the 16 critical and stationary points of the EPRL-FK amplitude can be obtained by computing the $4d$ bivectors at that point. Let $(g_a^{(+)},g_{a}^{(-)}$ be the pair of solutions from the $+$ and the $-$ sector, then the bivector $B_{ab}$ associated to the link between the nodes $a$ and $b$ in the boundary graph is given by
\begin{align}\label{Eq:Isometry01}
B_{ab}\;\simeq\;\left(\vec{b}_{ab}^{(+)},\,\vec{b}^{(-)}_{ab}\right)\;=\;j_{ab}\left(g^{(+)}_a\vec{n}_{ab},\,g^{(-)}_a\vec{n}_{ab}\right).
\end{align}

\noindent To interpret these geometries, let us compute the 4-volume for each of these configurations, using (\ref{Eq:VolumeFormula}). Using the isomorphism between $\mathbb{R}^4\wedge \mathbb{R}^4\simeq \mathbb{R}^3\oplus\mathbb{R}^3$, one has, for two bivectors $B_1\simeq(\vec{b}_1^{(+)},\vec{b}_1^{(-)})$ and $B_2\simeq(\vec{b}_2^{(+)},\vec{b}_2^{(-)})$, that
\begin{align}\label{Eq:VolumeInTermsOf3dVectors}
*\left(B_1\wedge B_2\right)\;=\;\left(\vec{b}_1^{(+)}\cdot\vec{b}_2^{(+)}\,-\,
\vec{b}_1^{(-)}\cdot\vec{b}_2^{(-)}
\right).
\end{align}

\noindent With (\ref{Eq:Isometry01}) and (\ref{Eq:VolumeInTermsOf3dVectors}) one can go through all 16 pairs of solutions in table \ref{Tab:Table02}, and compute the contributions for all crossings in the boundary graph $\Gamma$. With these it is a straightforward exercise to compute all sixteen volumes $V_i$ associated to the 16 bivector geometries. The results are given in table \ref{Tab:Table03}.

\begin{figure}
\begin{center}
\begin{tabular}{c|l|c|l}
crit/stat pt & 4-volume $V$ & crit/stat pt & 4-volume $V$  \\[5pt] \hline
$(\Sigma_1,\Sigma_1)$ & $0$ & $(\Sigma_3,\Sigma_1)$ & $-V_1$  \\[5pt]
$(\Sigma_1,\Sigma_2)$ & $V_2$ & $(\Sigma_3,\Sigma_2)$ & $V_2-V_1$  \\[5pt]
$(\Sigma_1,\Sigma_3)$ & $V_1$  & $(\Sigma_3,\Sigma_3)$ & $0$  \\[5pt]
$(\Sigma_1,\Sigma_4)$ & $V_1+V_2$ & $(\Sigma_3,\Sigma_4)$ & $V_2$  \\[5pt]
$(\Sigma_2,\Sigma_1)$ & $-V_2$ & $(\Sigma_4,\Sigma_1)$ & $-V_1-V_2$  \\[5pt]
$(\Sigma_2,\Sigma_2)$ & $0$ & $(\Sigma_4,\Sigma_2)$ & $-V_1$ \\[5pt]
$(\Sigma_2,\Sigma_3)$ & $V_1-V_2$ & $(\Sigma_4,\Sigma_3)$ & $-V_2$  \\[5pt]
$(\Sigma_2,\Sigma_4)$ & $V_1$ & $(\Sigma_4,\Sigma_4)$ & $0$ \\[5pt]
\end{tabular}
\caption{The 4-volumes associated to the 16 critical and stationary points for the EPRL-FK model, with $V_1=j^2$ and $V_2=\frac{jk}{2}\sin\theta\sin\phi$.}\label{Tab:Table03}
\end{center}
\end{figure}

\noindent As one can see, there are four bivector geometries which have zero volume. These correspond to geometries in which the whole polytope lies degenerately in a 3d-hyperplane, with dihedral angles either $0$ or $\pi$. In the literature, the amplitudes for those have been dubbed ``weird terms''. 

In the Euclidean 4-simplex amplitude, the non-weird terms correspond to geometrical polytopes. In particular, the polytope appears twice, one for each orientation of space-time. The opposite orientation leads to a negative sign for the action, which results in the so-called ``cosine problem''. This is to say that the amplitude does not only contain $e^{i S}$, but also $e^{-iS}$, adding up to the cosine.

Also for the polytope we consider, we can see that for every term, there is one with the opposite sign, which corresponds to the same geometry, just with opposite orientation. The respective volumes are negatives of one another. One can see that there are two different geometries (and their respective negatives), which come with the two volumes
\begin{align*}
V_1+V_2\;&=\;j^2\,+\,\frac{jk}{2}\sin\theta\sin\phi\\[5pt]
V_1-V_2\;&=\;j^2\,-\,\frac{jk}{2}\sin\theta\sin\phi
\end{align*}

\begin{figure}[hbt!]
\begin{center}
\includegraphics[scale=0.5]{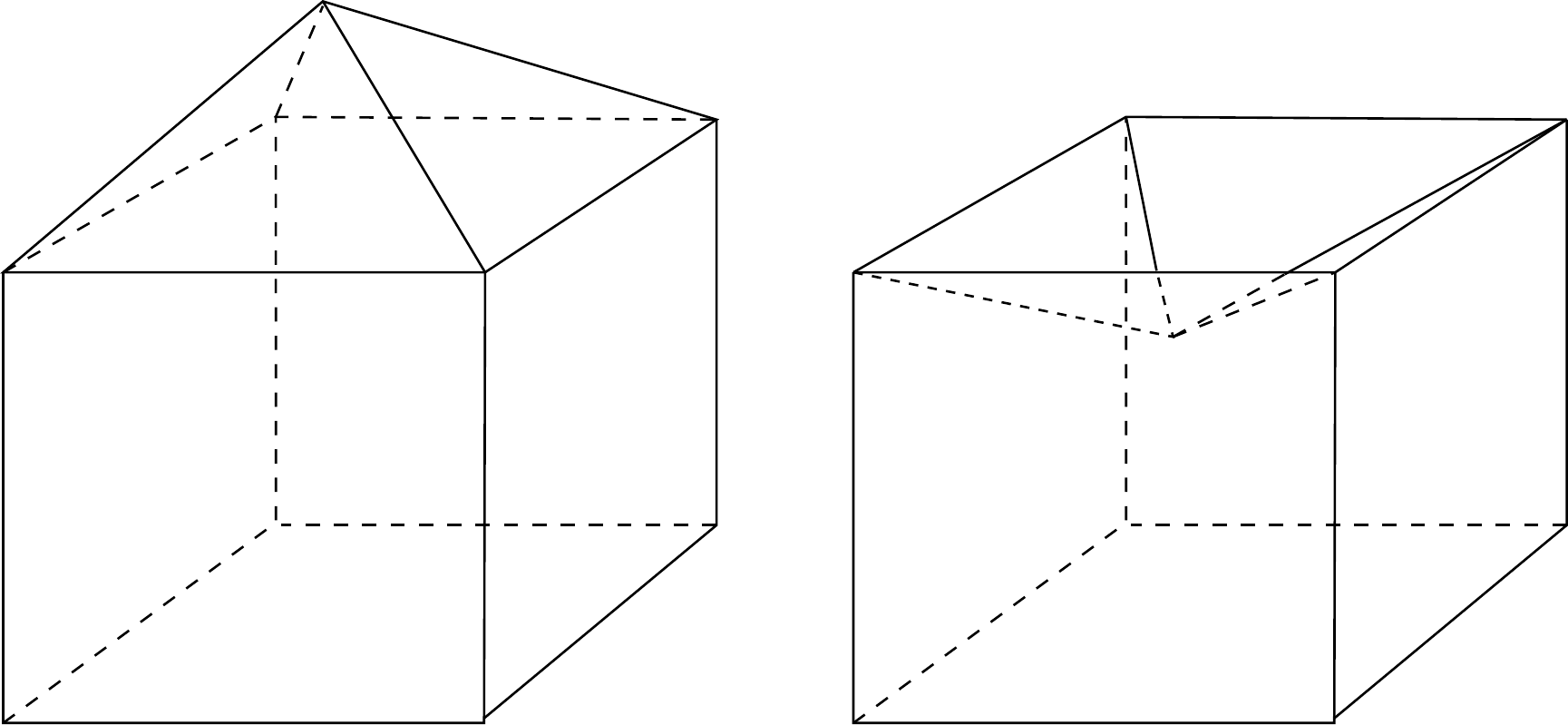}
\end{center}
\caption{($3d$ analog of) the two $4d$ geometries corresponding to the solutions $(\Sigma_1,\Sigma_4)$ and $(\Sigma_2,\Sigma_3)$. In one case, the ``roof'' of the house is pointing inwards, making this a non-convex $4d$ polytope.}\label{Fig:figure_099}
\end{figure}

\noindent By carefully considering the bivectors of these geometries, one can see that the first corresponds to a convex polytope, while the second one describes a non-convex polytope. Their 3d analogues are depicted in figure \ref{Fig:figure_099}. The reason for the presence of these two geometries can easily be understood from (\ref{Eq:CritStatEqns_OuterOuter} -- \ref{Eq:CritStatEqns_InnerInner}). The critical stationary equations for the EPRL-FK model are conditions for $SU(2)\times SU(2)$-rotations, to rotate 3d polytopes (given by the boundary data) such that they form a $4d$ geometry\footnote{More precisely, that the face bivectors of opposite rotated 3d polytopes are parallel, and the areas coincide. This can still leave room for non-metric configurations which manifest themselves as twisted or conformal geometries, see also \cite{Freidel:2010aq, Freidel:2013bfa}. The boundary data we have specified in (\ref{Eq:NewData}) excludes these, such that we get actual 4d polytopes.}. The boundary data (\ref{Eq:NewData}) actually admits two distinct 4d polytopes, one convex and one non-convex. 

There are still 8 critical / stationary points left over which are neither geometric polytopes, nor completely degenerate. In fact, a careful analysis reveals that they are part-polytope and part-degenerate. One can expect these hybrid-geometries to not appear in the Lorentzian signature model, while the non-convex ones are still likely to show up.

It should be clear from our construction that one can find more complex boundary graphs with boundary data which allow for an arbitrarily high number of critical stationary points! One simply has to take an existing set of boundary data, subdivide one of the 3-faces, and deform the new normals and spins slightly. This always doubles the solutions for the critical and stationary point in the asymptotic analysis of the EPRL-FK model. It is even apparent that one can thus construct boundary data, for which the amplitude has, in its asymptotic regime, geometries that correspond to polytopes with self-intersection. We could have already achieved this with our data (\ref{Eq:NewData}), had we chosen $k$ very large. 

\section{Summary and Discussion}\label{Sec:Summary}

In this article we considered non-convex polytopes in $\mathbb{R}^4$. In the first part, we have proven that a formula presented in \cite{Bahr:2018pfj} for the volume of convex polytopes also holds for the non-convex case. Here the main part was the careful definition of what we mean by non-convex polytope in this context. Although our definition is quite general, it seems plausible that the formula can be extended to include even more general cases, in particular those with self-intersections or those with a different boundary geometry. These generalisations will be left for future research.

In the second part we have considered non-convex polytopes in the EPRL-FK spin foam model. Moreover, we have demonstrated that, in general, boundary data for this model can lead to arbitrarily many geometrically distinct critical stationary points in the asymptotic analysis. The geometric reason for this is that certain $3d$ boundary data can be assembled in different ways to form the boundaries of different 4d polytopes (see e.g.~figure \ref{Fig:figure_099}). Of these, some will be non-convex, which could be demonstrated by using the volume formula from the first part. 

This is an extension of the results in \cite{Dona:2017dvf}, where the critical and stationary points of certain boundary graphs had been investigated. In particular, there the so-called ``twisted spikes'' were considered, which are graphs which have a node $n$ such that all other nodes of $\Gamma$ are connected to $n$. There the authors found that there are at most two solutions to the critical and stationary equations (\ref{Eq:CritStatEqns}) in the respective $SU(2)$ $BF$ theory. In our example, we found that there can be arbitrarily many solutions, if the boundary graph is sufficiently complex. In particular, in the boundary graph in figure \ref{Fig:figure_06a}, there are nodes which have neighbouring nodes which can be separated into non-connected and independent sets. This allows for independent solutions of the deficit angle $\theta$, leading to more than two solutions of (\ref{Eq:CritStatEqns}), resulting in non-convex polytopes in the EPRL-FK model.

It is an interesting question for future research how to avoid unwanted configurations, such as the non-convex ones\footnote{And indeed, whether they are unwanted or not. One might argue that diffeomorphism symmetry manifests itself in terms of vertex displacement symmetry, which should naturally lead to non-convex polytopes as gauge-equivalent to convex ones \cite{FreidelLouapreDiffeo2002, Dittrich:2008pw, Bahr:2009qc,Bahr:2009ku, Bahr:2011xs, Dittrich:2012qb}.}. One way might be an alteration of the model which includes the volume simplicity constraint in its linearised version. It is assumed that this would correspond to a closure constraint for the 4-normals, which, by Minkowski's theorem would only allow convex polytopes. Another might be a change in boundary data, i.e.~coherent states which do not only contain information about the 3-geometry of the boundary, but also its extrinsic curvature. We leave these points for future investigations. 

\section*{Acknowledgements}
This work was funded by the projects BA 4966/1-1 and BA4966/1-2 of the German Research Foundation (DFG). The authors would like to thank Pietro Dona and Simone Speziale for discussions.

%\cite{Abbott:2017xlt}

\bibliography{VolumeBib}
\bibliographystyle{ieeetr}

\end{document}